\documentclass[12pt, letterpaper]{article} 
\makeatletter\if@twocolumn\PassOptionsToPackage{switch}{lineno}\else\fi\makeatother
\usepackage[margin=1in,footskip=0.25in]{geometry}

\usepackage{tabulary,graphicx,amsmath,amsfonts,amssymb}
\usepackage[utf8x]{inputenc}
\usepackage[T1]{fontenc}
\usepackage[numbers]{natbib}
\usepackage{cmbright}
\usepackage{setspace}
\usepackage{caption}

\usepackage{authblk}
\title{Detection of biological signals from a live mammalian muscle using a diamond quantum sensor}
\author[1]{James Luke Webb}
\affil[1]{Center for Macroscopic Quantum States (bigQ), Department of Physics, Technical University of Denmark, Kgs. Lyngby, Denmark}
\author[1]{Luca Troise}
\affil[2]{Department of Neuroscience, University of Copenhagen, Copenhagen, Denmark}
\author[2]{Nikolaj Winther Hansen}
\affil[3]{Department of Health Technology, Technical University of Denmark, Kgs. Lyngby, Denmark}
\author[3]{Christoffer Olsson}
\affil[4]{Jagiellonian University, Krakow, Poland}
\author[4]{Adam M. Wojciechowski}
\affil[5]{Laboratoire des Sciences des Proc\'ed\'es et des Mat\'eriaux, Universit\'e Sorbonne Paris Nord,  93430 Villetaneuse, France}
\author[5]{Jocelyn Achard}
\author[5]{Ovidiu Brinza}
\affil[6]{Division Applied Quantum System, Felix Bloch Institute for Solid State Physics, Leipzig University, 04103, Leipzig, Germany}
\affil[7]{Danish Research Centre for Magnetic Resonance, Centre for Functional and Diagnostic Imaging and Research, Copenhagen University Hospital Hvidovre, Denmark}
\author[6]{Robert Staacke}
\author[6]{Michael Kieschnick}
\author[6]{Jan Meijer}
\author[3,7]{Axel Thielscher}
\author[2]{Jean-Fran\c{c}ois Perrier}
\author[1]{Kirstine Berg-S{\o}rensen}
\author[1]{Alexander Huck}
\author[1]{Ulrik Lund Andersen}
\date{}                     %% if you don't need date to appear
\begin{document}
\onehalfspacing
\maketitle

\begin{abstract}
The ability to perform noninvasive, non-contact measurements of electric signals produced by action potentials is essential in biomedicine. A key method to do this is to remotely sense signals by the magnetic field they induce. Existing methods for magnetic field sensing of mammalian tissue, used in techniques such as magnetoencephalography of the brain, require cryogenically cooled superconducting detectors. These have many disadvantages in terms of high cost, flexibility and limited portability as well as poor spatial and temporal resolution. In this work we demonstrate an alternative technique for detecting magnetic fields generated by the current from action potentials in living tissue using nitrogen vacancy centres in diamond. With 50pT/$\sqrt{Hz}$ sensitivity, we show the first measurements of sensing from mammalian tissue with a diamond sensor using mouse muscle optogenetically activated with blue light. We show these measurements can be performed in an ordinary, unshielded lab environment and that the signal can be easily recovered by digital signal processing techniques.
\end{abstract}

\section{Introduction}

Sensing of signals produced by living tissue is an essential tool for both medical diagnostics and for advancing the fundamental understanding of the structure and functioning of biological systems. Such signals, generated by propagating action potentials, are of particular importance in excitable cells such as neurons and muscle cells, allowing the cell-to-cell communication and movement that is essential for the functioning of the tissue and the organism as a whole\citep{nerobook}. Action potential can be measured using electrical probes\citep{Scanziani2009}, but these are invasive and can only be positioned to give limited spatial resolution. Magnetic field sensing provides a route towards noninvasive, high resolution, high speed sensing. To date, techniques for sensing the biological magnetic fields have been primarily based on superconducting quantum interference devices (SQUIDs)\citep{Jak1964, Clarke2018,RevModPhys.65.413,Cohen652}.This approach requires bulky magnetic shielding and cryogenic cooling, thus preventing proximity studies of living tissue and delivering poor spatial resolution. 

Noninvasive stimulation and high-resolution imaging of magnetic fields in an unshielded, ambient environment can be realized by using nitrogen vacancy (NV) centres in diamond for magnetic field sensing\citep{Gruber1997,Doherty2013,Taylor2008}. NV centers are quantum defects that can provide broadband vector magnetic field sensing\citep{Schloss2018,Fescenko2020,Wolf2015,Chatzidrosos2017} and imaging with high spatial resolution under ambient conditions using the technique of optically detected magnetic resonance (ODMR)\citep{Delaney2010, Wojciechowski2018}. It has broad applicability in life science\citep{Schirhagl2014,Wu2016} particularly due to the high biocompatibility of diamond, which can be placed in contact or even within biological specimens\cite{McGuinness2011,Kucsko2013}.Thus far NV sensing has focused on static or slow processes, such as imaging magnetotactic bacteria\citep{LeSage2013,Davis2018}. As yet there has been limited demonstration of sensing biological electrophysiological signals via magnetic field using diamond, with the most notable work being that by Barry et al.\citep{Barry2016} for invertebrates. Difficulties have included reaching sufficient sensitivity, keeping the sample alive and undamaged during measurement, interference from stimulation artifacts and the presence of background magnetic noise. 

In this work, we report the first use of a diamond quantum sensor to measure action potentials \textit{in vitro} from a live mammalian specimen via their magnetic field.  We detect the induced field from the dissected leg muscle of a genetically modified mouse, using optogenetic stimulation of channelrhodopsin to induce the action potential through blue light stimulation. We achieve a magnetic field sensitivity of 50pT/$\sqrt{Hz}$ and demonstrate methods that allow the specimen to remain alive under stimulation for up to 20 hours. By using advanced data post-processing and filtering, we are able to demonstrate the first example of sensing of magnetic field from optogenetic stimulation of a biological system under ambient conditions in a noisy, unshielded laboratory. We consider these measurements an important step towards the goal of \textit{in vivo} biosensing from living specimens, with the particular end goal of demonstrating sensing from neural networks in the mammalian brain\citep{Hall2012,Karadas2018} . 

\section{Results}

We used an inverted microscope containing a diamond magnetic field sensor, consisting of a single crystalline diamond sample with a 20$\mu$m layer comprising a high density of NV centers at the top facet (Figure \ref{fig:setup}, see Methods section). The biological specimen was placed near the NV surface separated only by a foil/insulator layer, thereby ensuring high proximity of the specimen to the sensing NV layer. Laser light at 532nm and frequency swept microwaves were applied to the NV sensor, while the induced fluorescence from the NV centers was imaged onto a photodetector. The time-varying magnetic field from the specimen was then detected using the protocol of optically detected magnetic resonance (ODMR) magnetometry. 

\subsection{Dynamic Range and Background Noise}
The diamond sensor was capable of measuring all the ambient background noise up to the kHz frequency range while maintaining maximum sensitivity without the sensor signal output saturating. An example of the raw magnetometer signal measured can be seen in Figure \ref{fig:rawsig},a). Assuming an ODMR linewidth of 1MHz, the approximate dynamic range without loss of sensitivity was estimated as 42$\mu$T, comfortably above the 600nT level of the predominant 50Hz and 150Hz background noise. 

We first measured the background noise detected by the magnetometer with deionised water in the chamber but without a muscle. Figure \ref{fig:rawsig},b) shows the amplitude spectral density, measured at microwave frequency on resonance (magnetically sensitive). Our noise floor was approximately 50pT/$\sqrt{Hz}$, defined by contributions from the electronic noise of the amplifiers and photodetector and the shot noise of the detected fluorescence. Here the shot noise limited sensitivity was approximately 8pT/$\sqrt{Hz}$. To characterise the noise, we measured the magnetometer output over many hours. The result can be seen in the spectrogram in Figure \ref{fig:rawsig},c), showing the range ($<$1kHz) where we expected to observe a biological signal. The two largest noise peaks can be seen at 50Hz and 150Hz, as expected from magnetic field detection of the fundamental mains frequency and from field produced by equipment transformers, each peak broadened by variable phase drift. Aside from mains harmonics, we observe a number of broad and narrowband noise sources. The majority of these we attribute to variable load airconditioning and water pumps in the building where the experiment was located, including some equipment from United States manufacturers that produced 60Hz fields. We consider this background typical for a research lab or clinical environment. 

\subsection{Muscle Electrophysiology}
Figure \ref{fig:elsig},a) shows a sketch of the fundamental biological process to be measured, where stimulation with light triggers a cascading opening of ion channels, generating an action potential (producing current flow and magnetic field) along the muscle. Further details on this process are given in Supplementary Information. Prior to the magnetometry experiment, this optogenetic stimulation was tested in a preliminary investigation in a separate setup capable of measuring action potential and muscle extension force. An example of a stimulation, measuring action potential using electrical probes and by measuring the force resulting from subsequent muscle contraction can be seen in Figure \ref{fig:elsig},b) . This test setup was used to determine the intensity of light required for good stimulation. No saturation in the electrical probe signal was observed up to the maximum intensity the light source could deliver. 

Figure \ref{fig:elsig},c) shows the response measured using an electrical probe contact to a stimulated muscle in the magnetometer sample chamber. We measured both the diamond sensor output and the electrical probe contact simultaneously, to give a complete picture of the muscle behaviour. The maximum biological signal amplitude as measured by the electrical probes versus time is given in Figure \ref{fig:elsig},d). The signal strength decreased over time as the muscle fatigued. This meant that after a certain time, a maximum signal to noise ratio was reached where further averaging would not help resolve the biological signal in the magnetic data. To find this point, we calculated the signal to noise ratio of the signal as a function of number of iterations during postprocessing. The rate of fatigue varied between different muscles, ranging from 8 hours in Figure \ref{fig:elsig} up to 16-18h. 

\subsection{Filtering Process}
Figure \ref{fig:ASDvsuppercut},a) shows the amplitude spectral density from Fast Fourier transforming the electrical probe data. The majority of the signal can be found in a frequency range from DC up to hundreds of Hz (blue histogram plot), thus unfortunately overlapping with the majority of the background magnetic noise. We make the reasonable assumption that the magnetic readout resembles the electrical probe readout since they originate from the same biological process. Therefore to filter the magnetic data, we limited the bandwidth to the range in which we expect a signal, thereby rejecting the majority of the background noise. Postprocessing the data collected, we imposed a digital bandpass filter from f$_{low}$=20Hz to a range of upper cutoff frequencies to determine the minimum at which the filter begins to corrupt the electrical probe data. We chose an upper cutoff of f$_{up}$=1.5kHz, to include as many of the signal frequencies as possible. It can be seen clearly from the spectrum in Figure \ref{fig:ASDvsuppercut},a) that this was more than sufficient to resolve the signal while excluding a significant amount of background noise. 

%notch filtering process
In order to remove the background noise within the measurement bandwidth, we Fourier transformed each 60s iteration dataset, selectively applied frequency domain notch filters corresponding to the noise peaks and then inverse Fourier transformed the data to recover the a filtered version of the timeseries. Due to the frequency overlap between signal and noise, it was critical to remove only parts of the signal that met two strict conditions: 1) to be clearly defined as noise (peak sufficiently above the white noise floor) and 2) only at those frequencies that did not distort the sought biological signal (on applying the same filter to the electrical probe data). Meeting only condition 1 would minimise noise while also removing the sought biological signal, whereas meeting only condition 2 would artificially recover the biological signal in the magnetic data by selection. 

We met these conditions by using two threshold values. The first, n$_{th}$ we define as the multiple above the median spectral amplitude a peak must exceed to be classed as noise. To apply this, we divided the spectrum into 40Hz wide windows, taking the median in each window m$_v$ and removed only those frequencies in each window that peaked above m$_{v}$$\times$n$_{th}$. By windowing, we avoided an excessive biasing of the filtering towards lower frequencies, due to the background 1/f$^{x}$ spectral slope. The second threshold value, m$_{th}$, we define as the percentage change in electrical probe signal relative to the unfiltered signal over a 40ms window which starts at the time of stimulation (t=0). These methods are clarified further in Supplementary Information.

We first removed those frequency components with the largest spectral amplitude (most likely to be noise) and continued until the SNR for each 60sec iteration was maximised, requiring between 60-200 notched frequencies. In the Supplementary Information, we show how this process can be simplified by first removing the broad 50/150Hz mains noise through time domain filtering, flattening the spectrum in the $<$200Hz range\citep{webb2}. We note that excluding the noise at $<$5Hz due to laser power fluctuations, 84 percent of the magnetic noise (4.8kHz bandwidth) was confined to 50Hz and 150Hz harmonics. As an aside, we show in the Supplementary Information that it is possible to significantly reduce the noise and recover a signal through use of a (n x 50Hz) comb of fixed-width notch filters. This configuration could be easily implemented in hardware for a sensor device for practical applications. 

\subsection{Biological Signal via Magnetic Field}

The timeseries for Nx60sec iterations for both electrical probe and magnetic data was then averaged. Data was obtained separately from two muscles. The result can be seen in Figure \ref{fig:magandelfiltered},a) and b). We observe $\sqrt{N}$ scaling (Figure \ref{fig:magandelfiltered},e) reaching an ultimate (rms) noise level of 22pT for Muscle 1 and 16pT for Muscle 2. The improved sensitivity for Muscle 2 was obtained with 12 hours more measurement averaging. For the second muscle, 2,3-Butanedione monoxime was added to the solution bath in order to inhibit movement without affecting the action potential. For Muscle 1, this compound was absent. A signal was observed in the magnetic data for both muscles typical of an action potential propagating along the muscle. This signal was present with and without muscle inhibitor, ruling out the signal being an artifact arising from muscle motion. For Muscle 2, a signal to noise ratio of 1 was reached after 32 iterations (30x32 simulations, 36 minutes measurement time), defining SNR as the averaged signal strength divided by the standard deviation of the averaged background magnetic noise. We phenomenologically modeled the expected action potential magnetic signal, full details of which are given in the Supplementary Information. The model parameters were within the range provided by literature and yield good agreement to the experimental data\citep{Cannon1993, Nikolic2009,Yizhar2011}.

We note that for Muscle 1 the diamond was placed approximately 2mm$\pm$1mm closer to the stimulation position along the muscle length than the electrical probes. This gap was not present for Muscle 2. The biological signal in the magnetic data for Muscle 1 was therefore consistently observed 1.5ms$\pm$0.5ms ahead of the electrical probe readout. This gives a crude estimate of propagation velocity in the muscle of 0.5-3m/s\citep{Juel1988}. The observed delay rules out that the recorded magnetic signal could stem from crosstalk pickup from the simultaneously measured electric probe circuit.

The difference in the shape and magnitude of the electrical readout between Muscle 1 and Muscle 2 arises from differences in contact quality between the muscles and the silver chloride probe electrodes. As a result of a reduced contact quality to Muscle 2, the signal strength was lowered and additional capacitance was introduced leading to distortions of the signal probed by the electrodes. This effect is not present on the magnetometer readout where we saw a sharp response, thus representing an advantage of the magnetic sensing over conventional electrophysiology. 

\section{Discussion} 

Using a diamond quantum sensor with pT-scale sensitivity to magnetic field and kHz measurement bandwidth, this work provides the first demonstration of sensing of the magnetic field from a signal generated by a living, mammalian biological specimen. We show that the sample can be kept alive for many hours while being measured and that the signal resembles that typical of action potentials measured by conventional electrical probes, without the drawbacks of poor electrical contact adding capacitive distortion. We measure a time delay between magnetic and electrical probe signal consistent with signal propagation along the muscle. Using optogenetic activation and comparison to a muscle where motion had been inhibited ensured the signal we measure was free of artifacts. The magnetometry technique is not dependent on optogenetic stimulation and is widely applicable to conventional electrical probe stimulation, or where stimulation originates from the living specimen itself. 

Using digital signal processing techniques, we show that a weak magnetic signal can be recovered in a noisy background without magnetic shielding, even in an ordinary laboratory environment with a significant degree of background magnetic noise typical of that in a large, busy building at a university or a hospital. Unlike alternative methods for high-sensitivity magnetometry, the high dynamic range of the diamond sensor allows the background noise to be recorded without saturation. Since the sensor does not saturate, the background noise can be directly detected and can thus be removed by adaptive windowed notch filtering. We show that this can also be done to a reasonable degree using fixed-width notch filters at mains harmonics frequencies. This could be implemented in hardware for realtime filtering in a portable sensor device to be used in a research or clinical environment\citep{Webb2019}. Future advances in sensitivity will only help improve the clear identification of the different sources of background noise, and could eventually allow active cancellation of magnetic noise in a small volume using a second sensor. 

The capability of operating in an ordinary lab or clinical environment without relying on superconducting technology, would open the door to many new research and diagnostic possibilities. A number of competing technologies seek to do this, most notably atomic vapour magnetometers\citep{Boto2017, Boto2018,Jensen2018,Jensen2016}. Although they are thus far superior in sensitivity, compared to diamond NV sensing, they have a number of disadvantages such as lack of biocompatibility, low dynamic range, inability to perform vector sensing in a single sensor, the need to screen from the Earth's magnetic field to achieve maximum sensitivity and low bandwidth at maximum sensitivity ($<$100Hz for a recent commercial atomic vapour sensor from QuSpin, Inc.\citep{QZFM} used for biosensing) that can be insufficient to achieve the micro to few-millisecond (kHz) time resolution needed to sense many biological signals. 

Our results are an important proof of concept experiment towards the goal of sensing \textit{in vivo} signals from the mammalian brain. Our setup is designed to be capable of measuring signals from dissected brain tissue slices and such measurements will take place in the near future. Our setup is also designed to allow magnetic field and thus biological signal imaging. Karadas et al. \cite{Karadas2018} have calculated the level of magnetic field produced by typical neuronal signals in the hippocampus to be between 10pT and 1.5nT. This is within the sensitivity range for bulk sensing using our scheme, but further advancements in sensitivity are required for imaging. 

\section{Methods}
%MAX 3000 words

\subsection{Inverted Microscope}
Figure \ref{fig:setup},a) shows a simplified 3D schematic of our inverted microscope incorporating the diamond biosensor. For optical pumping, up to 2W of horizontally polarised 532nm green laser (Coherent Verdi G2) illumination could be delivered from below a raised platform at Brewster's angle for diamond (67 deg). Polarisation was controlled before incidence on the diamond using a half wave plate to ensure maximum power transmission. Red fluorescence from the diamond was collected separately from the incident green light using an aspheric, anti-reflective coated 12mm diameter condenser lens (Thorlabs ACL1210). Fluorescence light was directed onto an electronically balanced photodetector (New Focus Inc.). 6mW was the typical power of collected fluorescence for 2W of green laser light. A reference beam for the photodetector was obtained by splitting off a few-mW portion of the input beam using a polarising beamsplitter. 

\subsection{Diamond Preparation}
The diamond used in this work was a [100] oriented electronic-grade single crystal from Element Six with dimensions 2 x 2 x 0.5mm$^3$ overgrown by a 20$\mu$m thick nitrogen doped layer using chemical vapor deposition (CVD). Nitrogen content in the gas phase was optimised during the growth to reach a level of 5ppm of nitrogen-14. The diamond was then 2.25MeV proton irradiated with a fluence of 3x10$^{15}$ protons/cm$^{2}$ followed by annealing at 800$^{\circ}$C for 4 hours. This gave a NV$^{-}$ density ranging between 0.1 and 1 ppm. The diamond was mounted into a central hole of a laser cut aluminum nitrate heatsink plate of dimensions 3x3x0.05cm$^3$. We measured an ODMR linewidth of 1MHz with a contrast of approximately 1.5 percent for each nitrogen-14 hyperfine transition.  

\subsection{Sensor Geometry}

\begin{figure}[h]
    \centering
    \includegraphics[width=1.0\textwidth]{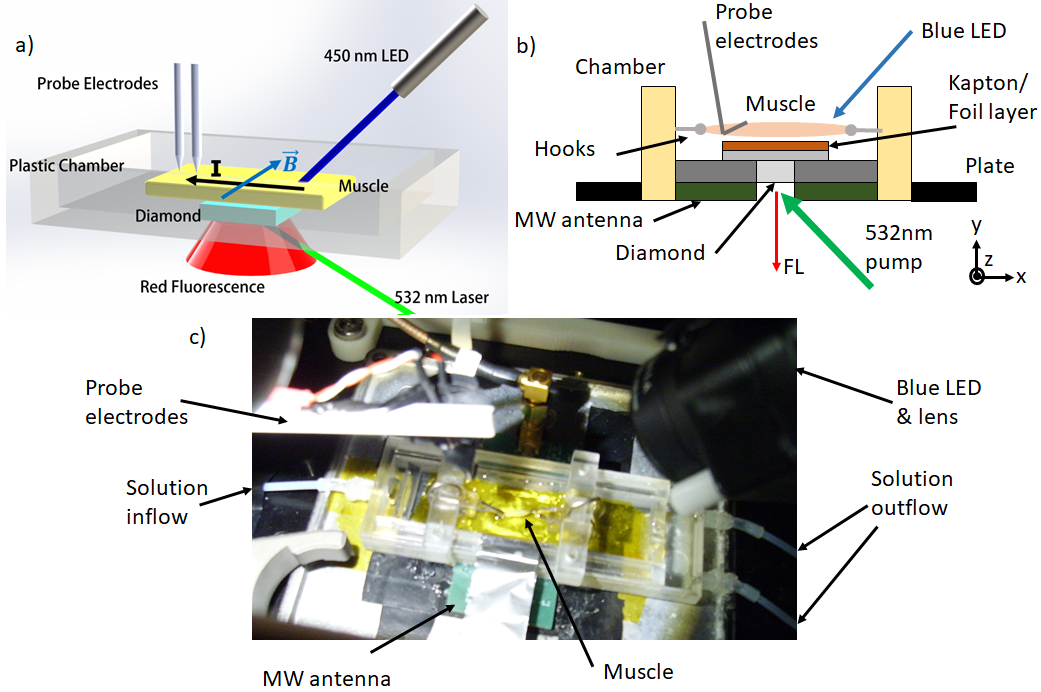}
    \caption{}
    \label{fig:setup}
\end{figure}

\begin{figure}[h]
    \centering
    \includegraphics[width=1.0\textwidth]{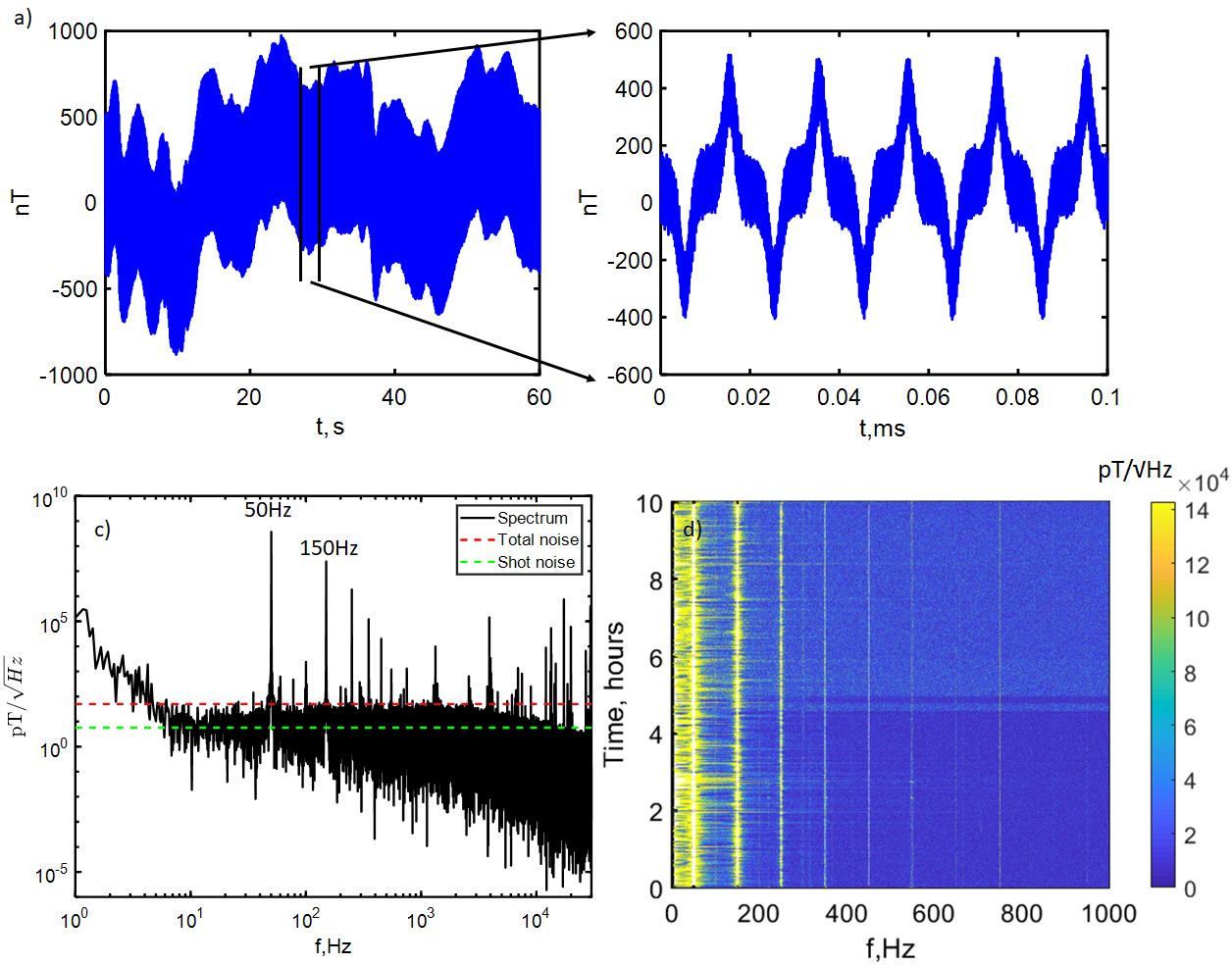}
    \caption{}
    \label{fig:rawsig}
\end{figure}

\begin{figure}[h]
    \centering
    \includegraphics[width=1.0\textwidth]{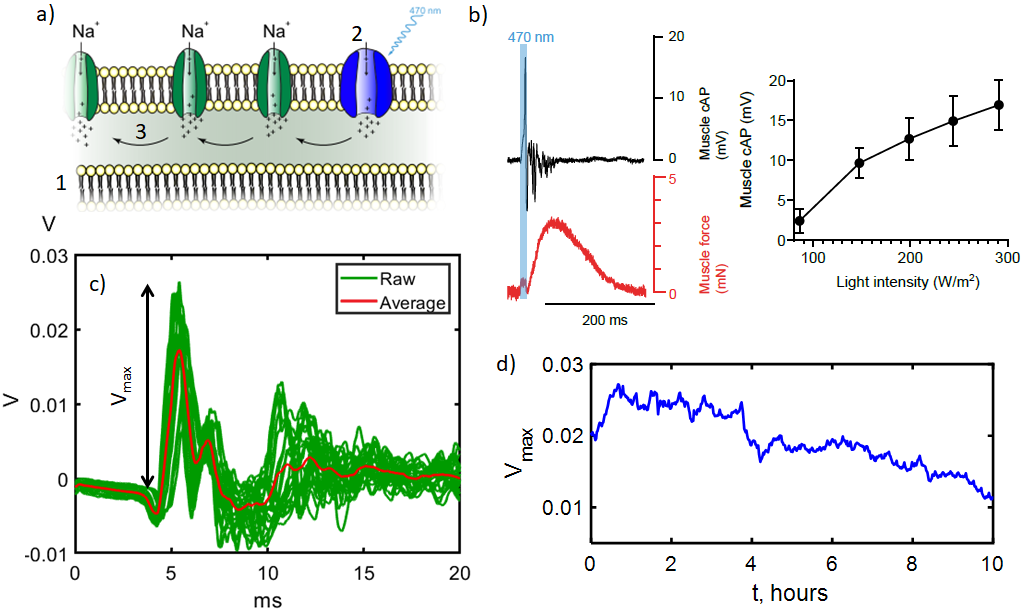}
    \caption{}
    \label{fig:elsig}
\end{figure}

\begin{figure}[h]
    \centering
    \includegraphics[width=1.0\textwidth]{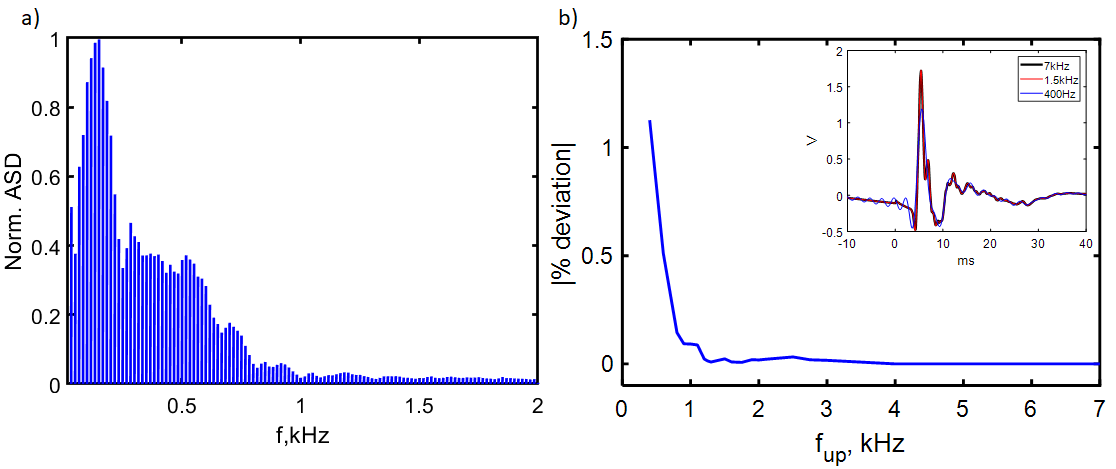}
    \caption{}
    \label{fig:ASDvsuppercut}
\end{figure}

\begin{figure}[h]
    \centering
    \includegraphics[width=0.8\textwidth]{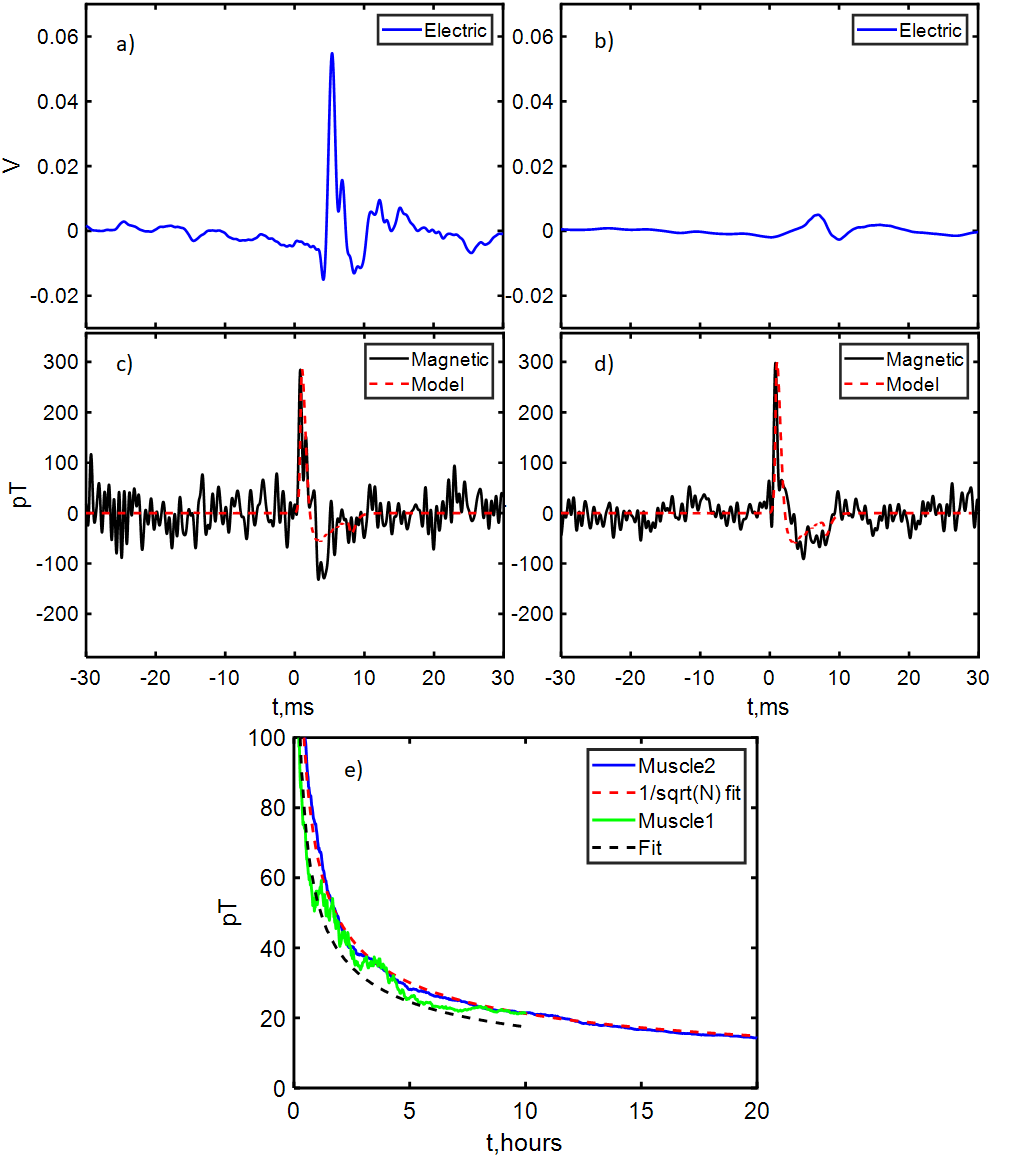}
    \caption{}
    \label{fig:magandelfiltered}
\end{figure}

The diamond and aluminium nitride plate were attached using watertight aquarium-safe silicone to a custom built broadband microwave antenna fabricated onto a printed circuit board with a hole for fluorescence collection from below (see schematic Figure \ref{fig:setup},b). On top of both antenna and plate was silicone mounted a rectangular 3D-printed, custom designed rectangular plastic sample chamber, which can be seen in Figure  \ref{fig:setup},c), that could hold a flow bath of solution, fed using a peristaltic pump. The chamber was fully accessible from above, allowing biological samples to be introduced and probe electrodes to contact the sample using micromanipulators. The sample was held on a pair of sliding hooks within the bath, directly above the top surface of the diamond. To protect the biological sample from laser heating, a 16$\mu$m thick layer of aluminum foil was placed on the top surface of the diamond, attached by 50$\mu$m Kapton tape in order to electrically insulate the foil and diamond from the sample. The resulting tens of micrometer separation between sample and diamond was undesirable due to reduction in magnetic field strength, but was taken as a precaution against sample heat damage based on previous experimental experience.

\subsection{Control and Readout}  

The microwave field was generated using a three-frequency drive scheme \citep{ElElla2017} using two radiofrequency (RF) generators (Stanford SG394) feeding a balanced mixer and then amplified (Minicircuits ZHL-16W-43+). One generator drove the transition between the $m_{s}=0$ and $m_{s}=\pm1$ of the ground state of the NVs with a frequency in the range of 2.7-3GHz and frequency modulated at 33kHz to implement lock-in detection. The second generator provided a fixed frequency of 2.16MHz to drive multiple hyperfine transitions. Two rare-earth magnets were aligned parallel to the (110) direction in the diamond and perpendicular to the main direction of signal current propagation, generating a DC bias field of $\sim$ 1.5mT. These directions are labelled on Figure \ref{fig:setup},a) and the field axis corresponds to ther z-axis on Figure \ref{fig:setup},b). We used a continuous wave (CW) scheme with constant microwave and laser power ensuring a stable (temperature) environment. Magnetometer sensitivity was optimised by adjusting the power of the reference beam to the balanced photodetector and by independently sweeping the power on the two RF signal generators to optimise microwave drive. 

Finally, the output voltage from the balanced detector was passed to a lock-in amplifier (Stanford SR850), from which the output was digitised by an analogue to digital converter (ADC, model NI PCI-6221) at 80kSa/sec. We term this channel the \textit{magnetic data}. We used a lock-in time constant of 30$\mu$s, giving a magnetic field measurement bandwidth of approximately 4.8kHz. The muscle was surface contacted by an electrical probe consisting of two L-shaped AgCl coated silver wires positioned 3 mm apart under the muscle mounted on a micromanipulator. The recording electrode voltage was amplified (Axon Cyberamp 320). This was then digitised at the same rate and simultaneously with the magnetic data. We term this channel the \textit{electrical probe data}.

\subsection{Specimen Preparation}

The muscle was stimulated optogenetically using blue light from a 470nm LED. Experiments were performed on genetically modified mice in which Channelrhodopsin 2 (ChR2), a light-gated cation channel, was used to create an action potential in the muscle. Animals were euthanized by cervical dislocation and extensor digitorum longus (EDL) muscles from both hind limbs were dissected in carbogen-saturated (95$\%$ O$_{2}$/5$\%$ CO$_2$) cold artificial cerebrospinal fluid (ACSF). Small suture loops were tied on distal and proximal tendons for later suspension in the recording chamber. Until use, EDL muscles were stored in a holding chamber continuously bubbled with carbogen. For some muscles, the myosin ATPase inhibitor 2,3-Butanedione monoxime (5mM in ACSF; Sigma) was added in order to uncouple excitation from contraction, ensuring that we measure only action potential and removing any possible artifacts arising from sample motion. Full details of the biological preparation are given in Supplementary Information with this work.

Prior to the experiment, the sample chamber and connecting tubing were cleaned by pumping heavily diluted household bleach through the system, followed by flushing with deionised water. This was then replaced with ACSF solution, carbogenated in a 500ml bottle and forming a closed circuit with the sample chamber. Temperature was measured in the chamber as 34$^{\circ}$C with laser and microwave power on. The mouse muscle was held in the chamber by suture loops on  hooks just above (but not in contact with) the diamond. 

\subsection{Stimulation Protocol}

The muscle was optically stimulated every 2sec, with a light pulse length of 5ms. The absence of contamination of the recording by a photovoltaic effect (Becquerel effect) induced by light was confirmed by taking traces recorded with the same protocols in the absence of muscle. Data from the magnetometer and from the electrodes was recorded from the ADC for 60sec data acquisition iterations during stimulation, giving 30 stimulations per iteration. The full data from both magnetic and electrical probe channels (2x 80kSa/sec x60sec) was stored for postprocessing. Many hours of data aquisition was possible. Postprocessing ensured that unexpected transient noise could be captured.  A random delay time was implemented between 60sec iterations (length between 10-30sec), ensuring each iteration began with a different mains phase to assist averaging. The absolute start time of each iteration was recorded and this measurement timeseries is used in the relevant plots in the results section. A fast ODMR sweep for selecting the optimal MW frequency was performed every five minutes during the measurement to compensate for any thermal drift.

%\section{Funding}
%The work presented here was funded by the Novo Nordisk foundation through the synergy grant \textit{bioQ} and the \textit{Center for Macroscopic States (bigQ)} funded by the Danish %National Research Foundation (DNRF).
%

\section{References}
\bibliographystyle{naturemag}
\bibliography{refspaper3mod}

\begin{thebibliography}{10}
\expandafter\ifx\csname url\endcsname\relax
  \def\url#1{\texttt{#1}}\fi
\expandafter\ifx\csname urlprefix\endcsname\relax\def\urlprefix{URL }\fi
\providecommand{\bibinfo}[2]{#2}
\providecommand{\eprint}[2][]{\url{#2}}

\bibitem{nerobook}
\bibinfo{author}{Kandel, E.~R.}, \bibinfo{author}{Schwartz, J.~H.},
  \bibinfo{author}{Jessell, T.~M.}, \bibinfo{author}{andA. J.~Hudspeth, S.
  A.~S.} \& \bibinfo{author}{Mack, S.}
\newblock \bibinfo{title}{Principles of neural science, fifth edition}
  (\bibinfo{year}{2013}).

\bibitem{Scanziani2009}
\bibinfo{author}{Scanziani, M.} \& \bibinfo{author}{H\"{a}usser, M.}
\newblock \bibinfo{title}{Electrophysiology in the age of light}.
\newblock \emph{\bibinfo{journal}{Nature}} \textbf{\bibinfo{volume}{461}},
  \bibinfo{pages}{930--939} (\bibinfo{year}{2009}).

\bibitem{Jak1964}
\bibinfo{author}{Jaklevic, R.~C.}, \bibinfo{author}{Lambe, J.},
  \bibinfo{author}{Silver, A.~H.} \& \bibinfo{author}{Mercereau, J.~E.}
\newblock \bibinfo{title}{Quantum interference effects in josephson tunneling}.
\newblock \emph{\bibinfo{journal}{Physical Review Letters}}
  \textbf{\bibinfo{volume}{12}}, \bibinfo{pages}{159--160}
  (\bibinfo{year}{1964}).

\bibitem{Clarke2018}
\bibinfo{author}{Clarke, J.}, \bibinfo{author}{Lee, Y.-H.} \&
  \bibinfo{author}{Schneiderman, J.}
\newblock \bibinfo{title}{Focus on {SQUIDs} in biomagnetism}.
\newblock \emph{\bibinfo{journal}{Superconductor Science and Technology}}
  \textbf{\bibinfo{volume}{31}}, \bibinfo{pages}{080201}
  (\bibinfo{year}{2018}).

\bibitem{RevModPhys.65.413}
\bibinfo{author}{H\"am\"al\"ainen, M.}, \bibinfo{author}{Hari, R.},
  \bibinfo{author}{Ilmoniemi, R.~J.}, \bibinfo{author}{Knuutila, J.} \&
  \bibinfo{author}{Lounasmaa, O.~V.}
\newblock \bibinfo{title}{Magnetoencephalography---theory, instrumentation, and
  applications to noninvasive studies of the working human brain}.
\newblock \emph{\bibinfo{journal}{Rev. Mod. Phys.}}
  \textbf{\bibinfo{volume}{65}}, \bibinfo{pages}{413--497}
  (\bibinfo{year}{1993}).
\newblock \urlprefix\url{https://link.aps.org/doi/10.1103/RevModPhys.65.413}.

\bibitem{Cohen652}
\bibinfo{author}{Cohen, D.}
\newblock \bibinfo{title}{Magnetic fields around the torso: Production by
  electrical activity of the human heart}.
\newblock \emph{\bibinfo{journal}{Science}} \textbf{\bibinfo{volume}{156}},
  \bibinfo{pages}{652--654} (\bibinfo{year}{1967}).

\bibitem{Gruber1997}
\bibinfo{author}{Gruber, A.}
\newblock \bibinfo{title}{Scanning confocal optical microscopy and magnetic
  resonance on single defect centers}.
\newblock \emph{\bibinfo{journal}{Science}} \textbf{\bibinfo{volume}{276}},
  \bibinfo{pages}{2012--2014} (\bibinfo{year}{1997}).

\bibitem{Doherty2013}
\bibinfo{author}{Doherty, M.~W.} \emph{et~al.}
\newblock \bibinfo{title}{The nitrogen-vacancy colour centre in diamond}.
\newblock \emph{\bibinfo{journal}{Physics Reports}}
  \textbf{\bibinfo{volume}{528}}, \bibinfo{pages}{1--45}
  (\bibinfo{year}{2013}).

\bibitem{Taylor2008}
\bibinfo{author}{Taylor, J.~M.} \emph{et~al.}
\newblock \bibinfo{title}{High-sensitivity diamond magnetometer with nanoscale
  resolution}.
\newblock \emph{\bibinfo{journal}{Nature Physics}}
  \textbf{\bibinfo{volume}{4}}, \bibinfo{pages}{810--816}
  (\bibinfo{year}{2008}).

\bibitem{Schloss2018}
\bibinfo{author}{Schloss, J.~M.}, \bibinfo{author}{Barry, J.~F.},
  \bibinfo{author}{Turner, M.~J.} \& \bibinfo{author}{Walsworth, R.~L.}
\newblock \bibinfo{title}{Simultaneous broadband vector magnetometry using
  solid-state spins}.
\newblock \emph{\bibinfo{journal}{Physical Review Applied}}
  \textbf{\bibinfo{volume}{10}} (\bibinfo{year}{2018}).

\bibitem{Fescenko2020}
\bibinfo{author}{Fescenko, I.} \emph{et~al.}
\newblock \bibinfo{title}{Diamond magnetometer enhanced by ferrite flux
  concentrators}.
\newblock \emph{\bibinfo{journal}{Physical Review Research}}
  \textbf{\bibinfo{volume}{2}} (\bibinfo{year}{2020}).

\bibitem{Wolf2015}
\bibinfo{author}{Wolf, T.} \emph{et~al.}
\newblock \bibinfo{title}{Subpicotesla diamond magnetometry}.
\newblock \emph{\bibinfo{journal}{Physical Review X}}
  \textbf{\bibinfo{volume}{5}} (\bibinfo{year}{2015}).

\bibitem{Chatzidrosos2017}
\bibinfo{author}{Chatzidrosos, G.} \emph{et~al.}
\newblock \bibinfo{title}{Miniature cavity-enhanced diamond magnetometer}.
\newblock \emph{\bibinfo{journal}{Physical Review Applied}}
  \textbf{\bibinfo{volume}{8}} (\bibinfo{year}{2017}).

\bibitem{Delaney2010}
\bibinfo{author}{Delaney, P.}, \bibinfo{author}{Greer, J.~C.} \&
  \bibinfo{author}{Larsson, J.~A.}
\newblock \bibinfo{title}{Spin-polarization mechanisms of the nitrogen-vacancy
  center in diamond}.
\newblock \emph{\bibinfo{journal}{Nano Letters}} \textbf{\bibinfo{volume}{10}},
  \bibinfo{pages}{610--614} (\bibinfo{year}{2010}).

\bibitem{Wojciechowski2018}
\bibinfo{author}{Wojciechowski, A.~M.} \emph{et~al.}
\newblock \bibinfo{title}{Precision temperature sensing in the presence of
  magnetic field noise and vice-versa using nitrogen-vacancy centers in
  diamond}.
\newblock \emph{\bibinfo{journal}{Applied Physics Letters}}
  \textbf{\bibinfo{volume}{113}}, \bibinfo{pages}{013502}
  (\bibinfo{year}{2018}).

\bibitem{Schirhagl2014}
\bibinfo{author}{Schirhagl, R.}, \bibinfo{author}{Chang, K.},
  \bibinfo{author}{Loretz, M.} \& \bibinfo{author}{Degen, C.~L.}
\newblock \bibinfo{title}{Nitrogen-vacancy centers in diamond: Nanoscale
  sensors for physics and biology}.
\newblock \emph{\bibinfo{journal}{Annual Review of Physical Chemistry}}
  \textbf{\bibinfo{volume}{65}}, \bibinfo{pages}{83--105}
  (\bibinfo{year}{2014}).

\bibitem{Wu2016}
\bibinfo{author}{Wu, Y.}, \bibinfo{author}{Jelezko, F.},
  \bibinfo{author}{Plenio, M.~B.} \& \bibinfo{author}{Weil, T.}
\newblock \bibinfo{title}{Diamond quantum devices in biology}.
\newblock \emph{\bibinfo{journal}{Angewandte Chemie International Edition}}
  \textbf{\bibinfo{volume}{55}}, \bibinfo{pages}{6586--6598}
  (\bibinfo{year}{2016}).

\bibitem{McGuinness2011}
\bibinfo{author}{McGuinness, L.~P.} \emph{et~al.}
\newblock \bibinfo{title}{Quantum measurement and orientation tracking of
  fluorescent nanodiamonds inside living cells}.
\newblock \emph{\bibinfo{journal}{Nature Nanotechnology}}
  \textbf{\bibinfo{volume}{6}}, \bibinfo{pages}{358--363}
  (\bibinfo{year}{2011}).

\bibitem{Kucsko2013}
\bibinfo{author}{Kucsko, G.} \emph{et~al.}
\newblock \bibinfo{title}{Nanometre-scale thermometry in a living cell}.
\newblock \emph{\bibinfo{journal}{Nature}} \textbf{\bibinfo{volume}{500}},
  \bibinfo{pages}{54--58} (\bibinfo{year}{2013}).

\bibitem{LeSage2013}
\bibinfo{author}{Sage, D.~L.} \emph{et~al.}
\newblock \bibinfo{title}{Optical magnetic imaging of living cells}.
\newblock \emph{\bibinfo{journal}{Nature}} \textbf{\bibinfo{volume}{496}},
  \bibinfo{pages}{486--489} (\bibinfo{year}{2013}).

\bibitem{Davis2018}
\bibinfo{author}{Davis, H.~C.} \emph{et~al.}
\newblock \bibinfo{title}{Mapping the microscale origins of magnetic resonance
  image contrast with subcellular diamond magnetometry}.
\newblock \emph{\bibinfo{journal}{Nature Communications}}
  \textbf{\bibinfo{volume}{9}} (\bibinfo{year}{2018}).

\bibitem{Barry2016}
\bibinfo{author}{Barry, J.~F.} \emph{et~al.}
\newblock \bibinfo{title}{Optical magnetic detection of single-neuron action
  potentials using quantum defects in diamond}.
\newblock \emph{\bibinfo{journal}{Proceedings of the National Academy of
  Sciences}} \textbf{\bibinfo{volume}{113}}, \bibinfo{pages}{14133--14138}
  (\bibinfo{year}{2016}).

\bibitem{Hall2012}
\bibinfo{author}{Hall, L.~T.} \emph{et~al.}
\newblock \bibinfo{title}{High spatial and temporal resolution wide-field
  imaging of neuron activity using quantum {NV}-diamond}.
\newblock \emph{\bibinfo{journal}{Scientific Reports}}
  \textbf{\bibinfo{volume}{2}} (\bibinfo{year}{2012}).

\bibitem{Karadas2018}
\bibinfo{author}{Karadas, M.} \emph{et~al.}
\newblock \bibinfo{title}{Feasibility and resolution limits of opto-magnetic
  imaging of neural network activity in brain slices using color centers in
  diamond}.
\newblock \emph{\bibinfo{journal}{Scientific Reports}}
  \textbf{\bibinfo{volume}{8}} (\bibinfo{year}{2018}).

\bibitem{webb2}
\bibinfo{author}{Webb, J.} \emph{et~al.}
\newblock \bibinfo{title}{Sensing of magnetic fields from biological signals
  using diamond nitrogen vacancy centres}.
\newblock \emph{\bibinfo{journal}{arXiv}}  (\bibinfo{year}{2020}).

\bibitem{Cannon1993}
\bibinfo{author}{Cannon, S.}, \bibinfo{author}{Brown, R.} \&
  \bibinfo{author}{Corey, D.}
\newblock \bibinfo{title}{Theoretical reconstruction of myotonia and paralysis
  caused by incomplete inactivation of sodium channels}.
\newblock \emph{\bibinfo{journal}{Biophysical Journal}}
  \textbf{\bibinfo{volume}{65}}, \bibinfo{pages}{270--288}
  (\bibinfo{year}{1993}).

\bibitem{Nikolic2009}
\bibinfo{author}{Nikolic, K.} \emph{et~al.}
\newblock \bibinfo{title}{Photocycles of channelrhodopsin-2}.
\newblock \emph{\bibinfo{journal}{Photochemistry and Photobiology}}
  \textbf{\bibinfo{volume}{85}}, \bibinfo{pages}{400--411}
  (\bibinfo{year}{2009}).

\bibitem{Yizhar2011}
\bibinfo{author}{Yizhar, O.}, \bibinfo{author}{Fenno, L.~E.},
  \bibinfo{author}{Davidson, T.~J.}, \bibinfo{author}{Mogri, M.} \&
  \bibinfo{author}{Deisseroth, K.}
\newblock \bibinfo{title}{Optogenetics in neural systems}.
\newblock \emph{\bibinfo{journal}{Neuron}} \textbf{\bibinfo{volume}{71}},
  \bibinfo{pages}{9--34} (\bibinfo{year}{2011}).

\bibitem{Juel1988}
\bibinfo{author}{Juel, C.}
\newblock \bibinfo{title}{Muscle action potential propagation velocity changes
  during activity}.
\newblock \emph{\bibinfo{journal}{Muscle {\&} Nerve}}
  \textbf{\bibinfo{volume}{11}}, \bibinfo{pages}{714--719}
  (\bibinfo{year}{1988}).

\bibitem{Webb2019}
\bibinfo{author}{Webb, J.~L.} \emph{et~al.}
\newblock \bibinfo{title}{Nanotesla sensitivity magnetic field sensing using a
  compact diamond nitrogen-vacancy magnetometer}.
\newblock \emph{\bibinfo{journal}{Applied Physics Letters}}
  \textbf{\bibinfo{volume}{114}}, \bibinfo{pages}{231103}
  (\bibinfo{year}{2019}).

\bibitem{Boto2017}
\bibinfo{author}{Boto, E.} \emph{et~al.}
\newblock \bibinfo{title}{A new generation of magnetoencephalography: Room
  temperature measurements using optically-pumped magnetometers}.
\newblock \emph{\bibinfo{journal}{{NeuroImage}}}
  \textbf{\bibinfo{volume}{149}}, \bibinfo{pages}{404--414}
  (\bibinfo{year}{2017}).

\bibitem{Boto2018}
\bibinfo{author}{Boto, E.} \emph{et~al.}
\newblock \bibinfo{title}{Moving magnetoencephalography towards real-world
  applications with a wearable system}.
\newblock \emph{\bibinfo{journal}{Nature}} \textbf{\bibinfo{volume}{555}},
  \bibinfo{pages}{657--661} (\bibinfo{year}{2018}).

\bibitem{Jensen2018}
\bibinfo{author}{Jensen, K.} \emph{et~al.}
\newblock \bibinfo{title}{Magnetocardiography on an isolated animal heart with
  a room-temperature optically pumped magnetometer}.
\newblock \emph{\bibinfo{journal}{Scientific Reports}}
  \textbf{\bibinfo{volume}{8}} (\bibinfo{year}{2018}).

\bibitem{Jensen2016}
\bibinfo{author}{Jensen, K.} \emph{et~al.}
\newblock \bibinfo{title}{Non-invasive detection of animal nerve impulses with
  an atomic magnetometer operating near quantum limited sensitivity}.
\newblock \emph{\bibinfo{journal}{Scientific Reports}}
  \textbf{\bibinfo{volume}{6}} (\bibinfo{year}{2016}).

\bibitem{QZFM}
\bibinfo{author}{QuSpin}.
\newblock \bibinfo{title}{Technical documentation on {QZFM Gen-2 OPM} sensor,
  {QuSpin Inc}}  (\bibinfo{year}{2020}).

\bibitem{ElElla2017}
\bibinfo{author}{El-Ella, H. A.~R.}, \bibinfo{author}{Ahmadi, S.},
  \bibinfo{author}{Wojciechowski, A.~M.}, \bibinfo{author}{Huck, A.} \&
  \bibinfo{author}{Andersen, U.~L.}
\newblock \bibinfo{title}{Optimised frequency modulation for continuous-wave
  optical magnetic resonance sensing using nitrogen-vacancy ensembles}.
\newblock \emph{\bibinfo{journal}{Optics Express}}
  \textbf{\bibinfo{volume}{25}}, \bibinfo{pages}{14809} (\bibinfo{year}{2017}).

\end{thebibliography}
\section{Acknowledgments}
We would like to thank Carmelo Bellardita for helping us do the immunohistochemistry. We acknowledge the Core Facility for Integrated Microscopy, Faculty of Health and Medical Sciences, University of Copenhagen for using their confocal microscope for immunohistochemistry image acquisition. We acknowledge the assistance of Kristian Hagsted Rasmussen for fabrication and diamond processing and Mursel Karadas (former DTU Heath Technology, currently New York University) for contributions and prior experimental and theoretical modeling work. 

\section{Author contributions}
The project was conceived by AH and ULA. Methodology development, investigation and analysis were performed by JLW, LT, NWH and AMW. NWH performed all animal dissections. Modeling work was performed by CO. Diamond growth and irradiation was performed by JA, OB, RS, MK and JM. This manuscript was written by JLW with editing and review contribution by all other authors. The overall supervision was performed by AT, JFP, KBS, AH and ULA. 

\section{Competing Interests}
The authors declare that they have no known competing interests that would influence the work reported here. 

\pagebreak
\section{Figures and Legends}

\textbf{Figure 1: Experiment schematic and photograph.} a) Simplified 3D schematic of the magnetometer setup, showing the laser and blue LED illumination and fluorescence (FL) collection directions and sample chamber orientation, the direction of maximum magnetic field sensivity (B) and the direction of current flow (I) in the muscle.  b) Side view schematic (not to scale) of the chamber/diamond sensor/MW antenna stack, joined and affixed to a movable plate with silicone. c) Photograph from above of the chamber, showing solution inflow connections and the mouse muscle, below which the diamond lies separated by a layer of Kapton tape and aluminium foil acting as a heatsink.

\pagebreak

\textbf{Figure 2: Demonstration of high dynamic range, bandwidth and sensitivity to magnetic field.} a) (upper pane) The raw unfiltered magnetic signal for (left) a full 60sec iteration and (right) for a zoomed 0.1sec segment of the same iteration. The signal was dominated by low frequency and DC laser power drift ($<$5Hz), 50Hz and 150Hz noise from mains electricity and higher frequency ($>$10kHz) noise and ranges between $\pm$1$\mu$T, well within the dynamic range of the magnetometer. (lower pane) Spectral density in pT/$\sqrt{Hz}$ for b) a single 60sec iteration and c) a spectrogram of repeated 60sec acquisitions over 10 hours. The sensitivity floor is approximately 50pT/$\sqrt{Hz}$ with f(-3dB)=4.8Hz defined by the lock-in amplifier low pass filter. Also indicated are calculations of the total noise, which includes electronic and shot noise, and of the estimated shot noise level alone. Many sources of background magnetic noise can be seen to peak well above this floor.

\pagebreak

\textbf{Figure 3: Mouse muscle electrophysiology and signal variation over time} a) Sketch of the biological signal generation process. In the muscle cell bi-lipid membrane (1) channelrhodopsin (2) opening triggers influx of Na$^{+}$ ions (3), creating an action potential running along the muscle. b) Preliminary measurements taken on a separate setup of a single stimulation and readout via electrical probes (mV) and via muscle contraction force (mN). The strength of the signal as a function of light intensity is also shown. c) Example of the readout of the biological signal in the magnetometer setup from a muscle (Muscle 1) by the electrical contact probe. Here t=0ms is when the stimulation light is applied. The red trace shows the average signal observed over all stimulations. d) (left axis) Maximum size of the initial peak in the signal, which steadily drops by a factor of 2 over time as the muscle becomes fatigued.

\pagebreak

\textbf{Figure 4: Frequency spectrum of the biological signal and defining optimum filter thresholding.} a) Spectrogram of the normalised Fourier transform amplitudes of the electrical probe voltage data, showing that the majority of the signal frequency components (shaded blue region) are under 1.5kHz. b) Percentage deviation from the unfiltered signal as a function of upper bandpass cutoff frequency f$_{up}$. The signal begins to be significantly corrupted below 1.5kHz, as can be seen in the inset example for f$_{up}$=400Hz where t=0 is the stimulation time. 

\pagebreak

\textbf{Figure 5: Simultaneous electrical and magnetometer readout of the biological signal.} Magnetic and scaled electrical probe data for LED stimulation of two muscles a) left panes, Muscle 1: averaged for 8 hours (30x425 stimulations)  and b) right panes, Muscle 2: averaged for 16 hours (30x837 stimulations). The maximum signal strength was approximately 250pT. c) Noise on the filtered magnetic data as a function of time for Muscle 1 and Muscle 2, showing $\sqrt{N}$ dependence with the number of measurements taken (increasing time).

\end{document}